\documentclass[%
reprint,
 amsmath,amssymb,
 aps,
prb,
]{revtex4-1}

\usepackage{graphicx}
\usepackage{dcolumn}
\usepackage{bm}

\begin{document}
\preprint{APS/123-QED}

\title{Evidence for a New Excitation at the Interface Between a High-T$_c$ Superconductor and a Topological Insulator}

\author{Parisa Zareapour,$^{1}$ Alex Hayat,$^{1}$ Shu Yang F. Zhao,$^{1}$ Michael Kreshchuk,$^{1}$\\ Yong Kiat Lee,$^{1}$ Anjan A. Reijnders,$^{2}$ Achint Jain,$^{1}$ Zhijun Xu,$^{3}$ T. S. Liu,$^{3,4}$\\G.D. Gu,$^{3}$\ Shuang Jia,$^{5}$ Robert J. Cava ,$^{5}$and Kenneth S. Burch $^{1,6}$}
\affiliation{$^{1}$ Department of Physics and Institute for Optical Sciences, University of Toronto, 60 St George Street, Toronto, Ontario, Canada M5S 1A7.}%
\affiliation{$^{2}$ Montana Instruments, Bozeman, MT 59715.}%

\affiliation{$^{3}$ Department of Condensed Matter Physics and Materials Science (CMPMS), Brookhaven National Laboratory, Upton, New York 11973, USA.}%
\affiliation{$^{4}$ School of Chemical Engineering and Environment, North University of China, China.}%
\affiliation{$^{5}$ Department of Chemistry, Princeton University, Princeton, New Jersey 08544, USA.}%
\affiliation{$^{6}$ Department of Physics, Boston College, 140 Commonwealth Avenue, Chestnut Hill, MA 02467.}%

\date{\today}

\begin{abstract}
High-temperature superconductors exhibit a wide variety of novel excitations. If contacted with a topological insulator, the lifting of spin rotation symmetry in the surface states can lead to the emergence of unconventional superconductivity and novel particles. In pursuit of this possibility, we fabricated high critical-temperature (T$_c$ $\sim$ 85 K) superconductor/topological insulator (Bi$_2$Sr$_2$CaCu$_2$O$_{8+\delta}$/Bi$_2$Te$_2$Se) junctions. Below 75 K, a zero-bias conductance peak (ZBCP) emerges in the differential conductance spectra of this junction. The magnitude of the ZBCP is suppressed at the same rate for magnetic fields applied parallel or perpendicular to the junction. Furthermore, it can still be observed and does not split up to at least 8.5 T. The temperature and magnetic field dependence of the excitation we observe appears to fall outside the known paradigms for a ZBCP.

\end{abstract}

\maketitle

\section{\label{sec:level1}INTRODUCTION}

The observation of new excitations is central to our understanding of numerous physical phenomena from the Higgs Boson, the missing part of the standard model, to the Cooper pair and the collective phenomenon of superconductivity\cite{Wilczek:2009ek}. There has been growing interest in the excitations that occur at superconducting interfaces, such as Andreev bound states and Majorana fermions. Evidence of Majorana fermions was observed in superconductor/topological insulator (TI) interfaces and InSb nanowires\nocite{Anonymous:NKdIlZ7d,Zhang:2011bl,Veldhorst:2012gn,Rokhinson:2012epa,Das:2012hi,Deng:2012gn,Finck:2013ej}. Andreev bound states have been extensively studied and measured in superconducting josephson junctions and at the \{110\} surface of d-wave superconductors\cite{Deutscher:2005tm,1982PhRvB..25.4515B,Dirks:2011el,Aubin:2002go,Alff:1998td,Lee:2013gj}. In fact, Andreev bound states that emerge at the \{110\} surface of the cuprates are signatures of the unconventional superconducting ground state. Some recent theoretical proposals have suggested that in a topological d-wave superconductor, these Andreev bound states would be converted into Majorana Fermions\cite{Anonymous:H70z94jd}. Furthermore a number of theoretical proposals have pointed out the utility of using high-T$_c$ cuprates to induce novel superconducting states in topological insulators\cite{Anonymous:H70z94jd,Takei:2013dp,Kim:2012en,Lucignano:2012eu}.

More generally, the lifting of spin rotation symmetry in the surface states of a TI, suggests the superconducting proximity effect will be quite unconventional in these materials. Towards this goal, our group was the first to demonstrate a high-T$_c$ proximity effect in the topological insulators Bi$_2$Se$_3$ and Bi$_2$Te$_3$, using our new technique of Mechanical Bonding \cite{2012NatCo...3E1056Z}. However, the Fermi energy in the Bi$_2$Se$_3$ and Bi$_2$Te$_3$ was deep in the bulk conduction band due to large defect concentrations, limiting the ability to probe the surface states of the TI. An alternative material, Bi$_2$Te$_2$Se, is quite promising in this regard as the defect concentrations are known to be strongly suppressed. This has led to the observation of nearly insulating behaviour and strong suppression of bulk transport at low temperatures \cite{Xiong:2012ht,Neupane:2012ft,Ren:2010ji}. We have confirmed this in the Bi$_2$Te$_2$Se crystals used in this study via temperature dependent transport (Fig.S1 A). Interestingly, differential conductance measurements in some of our Bi$_2$Sr$_2$CaCu$_2$O$_{8+\delta}$ (Bi-2212)/Bi$_2$Te$_2$Se devices exhibit a zero-bias conductance peak (ZBCP) that behaves quite differently from previous experimental observations and theoretical predictions (Fig.\ref{fig:temp}A). Previous theoretical and experimental studies of various ZBCPs: Andreev bound states (ABS), coherent Andreev reflection (CAR), weak antilocalization (WAL), Andreev reflection, proximity effect, Kondo effect, magnetic impurities, and Majorana fermions have established that the ZBCP should split in applied magnetic field, appear at T$_c$, be broadened with temperature, and/or its height depends strongly on the orientation of the applied field \cite{Deutscher:2005tm,1982PhRvB..25.4515B,Giazotto:2001dq,2000Natur.405..764S,Tkachov:2005jt,Anonymous:H70z94jd}. The ZBCP we observe in mechanically-bonded Bi$_2$Sr$_2$CaCu$_2$O$_{8+\delta}$ (Bi-2212)/Bi$_2$Te$_2$Se devices are suppressed at the same rate in different magnetic field orientations(Fig. \ref{fig:mag}C), do not appear to broaden with applied fields or raised temperature(\ref{fig:width}), and are completely suppressed by $\sim$0.8T$_c$ (Fig. \ref{fig:temp}). Moreover, we can rule out heating effects, since the ZBCP is observed in a wide range of junction resistances  ($\sim0.1\rightarrow1~k\Omega$). Thus our results are completely inconsistent with previous observations or theoretical explanations of a ZBCP, suggesting a new effect emerging at the interface between a high-T$_c$ and a TI.

\section{\label{sec:level1}RESULTS AND DISCUSSION}

\begin{figure*}
\includegraphics[width=1\textwidth]{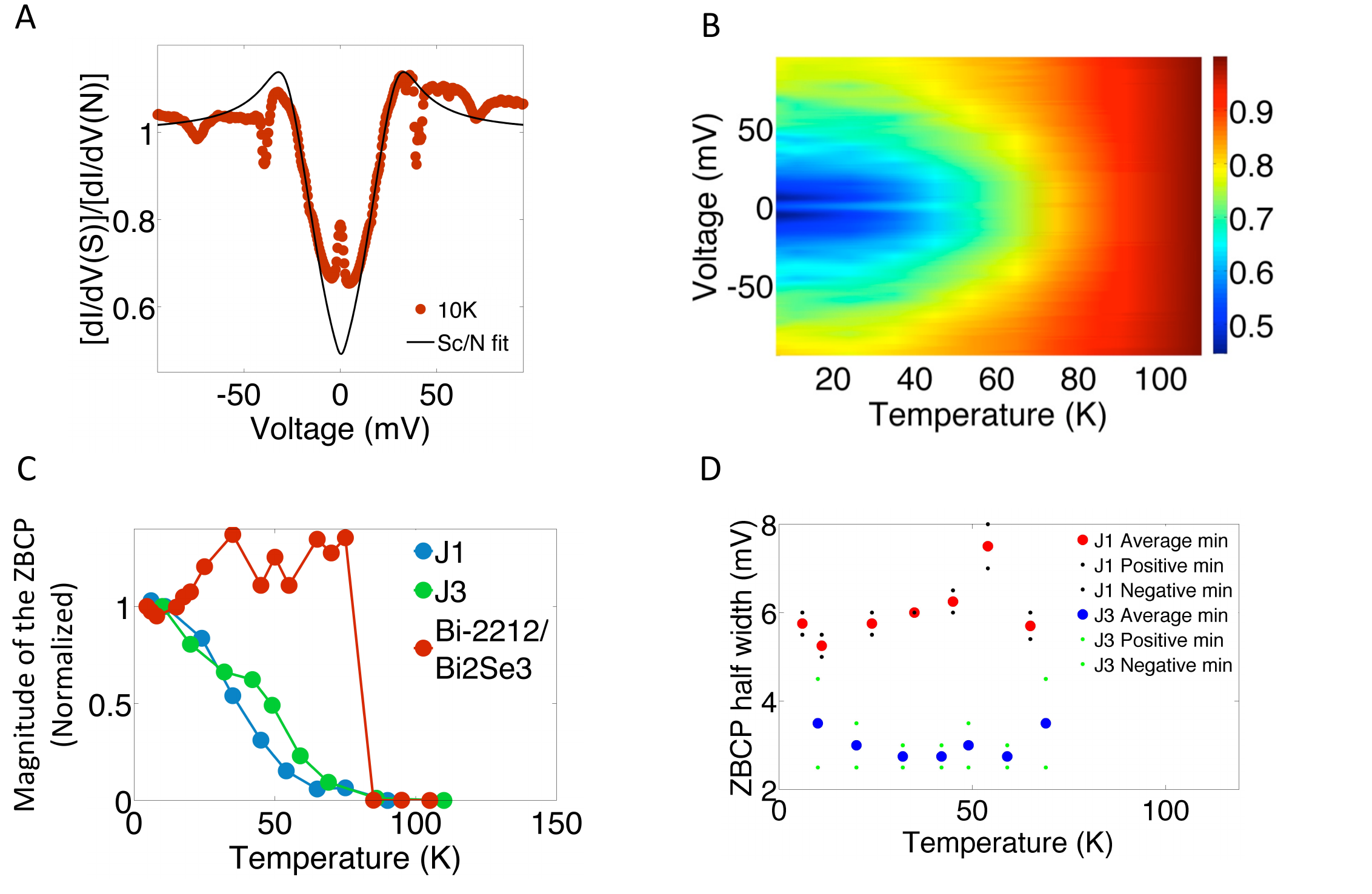}
\caption{\label{fig:temp} (color online) (\textbf{A}) Normalized differential conductance measurement of Bi-2212/Bi$_2$Te$_2$Se device 1 (J1), divided by the normal-state conductance. The black line shows the theoretical calculation for high-barrier d-wave superconductor/Normal tunnelling conductance for this device. The fitting parameters for J1 are $\Delta = 40 mV$, $\Gamma = 0.2 (\Delta$), and  $Z=0.8$. (\textbf{B}) Differential conductance characteristics of J1 divided by the normal-state conductance (110K) at different temperatures. (\textbf{C}) The temperature dependence of the ZBCP magnitude, normalized to its 10 K value, for the Bi-2212/Bi$_2$Te$_2$Se junctions 1 and 3, and a Bi-2212/Bi$_2$Se$_3$ proximity device. (\textbf{D}) The half width of the ZBCP in Bi-2212/Bi$_2$Te$_2$Se junctions 1 and 3, as a function of temperature, measured by finding the minimum of conductance at the positive voltage side and negative voltage side separately (small circles). The big circles show the average of the positive and negative minima. }
\end{figure*}

Low temperature ($\sim$ 10 K) dI/dV spectrum of a Bi-2212/Bi$_2$Te$_2$Se device is shown in Fig. \ref{fig:temp}A. Two main features are seen in this spectrum; a ZBCP, and an overall V-shaped background. This is expected since our cleaving method leaves Bi$_2$Te$_2$Se atomically flat over much smaller regions than Bi-2212.\cite{2012NatCo...3E1056Z} As such the overall device contains mostly high-barrier junctions, with a few areas in direct contact. Tunnelling from the much larger high-barrier regions results in a V-shaped dI/dV, typical for tunnelling into a d-wave superconductor. To confirm the origin of this background, we use an extension of the BTK theory for anisotropic superconductors, to calculate c-axis normal material/d-wave superconductor tunnelling conductance\cite{1982PhRvB..25.4515B,Kashiwaya:1996wz}. The black line in Fig. \ref{fig:temp}A shows the d-wave Superconductor (Sc)/Normal (N) conductance fit (see supplemental). This Sc/N background was observed in all Bi-2212/Bi$_2$Te$_2$Se junctions (Fig. S1 D). Interestingly, in some of the devices we observed a regular series of resonances in the conductance spectra originating from McMillan-Rowell oscillations, though their appearance was not correlated with the observation of the ZBCP (Fig. \ref{fig:temp}A, Fig. S2 A) \cite{Shkedy:2004kca}. Temperature dependence of the differential conductance spectra of Bi-2212/Bi$_2$Te$_2$Se device 1 (J1) is shown in Fig. \ref{fig:temp}B. For every temperature (T) shown, $dI/dV (T)$ curves are normalized to the normal-state conductance, taken at 110 K. As the temperature is lowered below the T$_c$ of Bi-2212, we observe the clear opening of the superconducting gap starting at T$_c$=85 K, consistent with our previous studies of high barrier junctions between Bi-2212 and a variety of materials (GaAs, Graphite) \cite{2012NatCo...3E1056Z,Hayat:2012jo} as well as other tunnelling measurements\cite{Pan:509340,Misra:2002it,2003PhyC..387..162G,2003PhyC..387..162G,Deutscher:2005tm,Vedeneev:2005gz,Kasai:2010kn,1991PhRvB..4311492W,Fischer:2007fa}. The conductance at high-bias (and the overall spectra) decreases continuously as the temperature is lowered, partially due to the Bi$_2$Te$_2$Se becoming more resistive (Fig. S1 A). (see supplemental) To remove any temperature dependence of the spectra not due to the ZBCP, we measured the strength of the ZBCP by taking the difference between the normalized dI/dV at zero bias and its minimum value (the cutoff voltage of the zero-bias peak). The amplitude decays in a manner similar to the closing of an order parameter (see Fig. \ref{fig:temp}C). To look for thermal broadening, we measured the width as the average of the positive and negative voltages of the conductance minima. As can be seen in Fig. \ref{fig:temp}D, the ZBCP does not decohere as the temperature increases, but stays constant within our experimental error.

This temperature dependence is inconsistent with ZBCPs emerging from a standard Andreev reflection\cite{DRHeslinga:2011wv} and/or proximity effect\cite{1982PhRvB..25.4515B}. For instance, in our previous work on Bi$_2$Se$_3$\cite{2012NatCo...3E1056Z}, the conductance at zero bias increased to twice the normal conductance right below T$_c$, in agreement with theory\cite{1982PhRvB..25.4515B}. This factor of two is expected at transparent interfaces, when the Andreev reflection happens at the interface, since the incoming electron forms a cooper pair and thus two electrons cross the interface, resulting in a doubling of conductance. However in less transparent interfaces, the conductance at zero bias is the first to be reduced and the shape is significantly altered. The conductance of the Bi-2212/Bi$_2$Te$_2$Se junction however, decreases with temperature continuously and the ZBCP starts to develop at $T_{ZBCP}\sim0.8T_c$ (Fig. \ref{fig:temp}C, Fig. S1 D). The emergence of the ZBCP well below T$_c$ (Fig. S1 D) would also appear to eliminate another possible explanation. As discussed earlier, one can observe an ABS by tunnelling into the AB-plane of the Bi-2212 (\{110\} surface). However, previous studies have found that these states will be observed at T$_c$, not well below it.\cite{Deutscher:2005tm,Aubin:2002go}

\begin{figure*}
\includegraphics[width=1\textwidth]{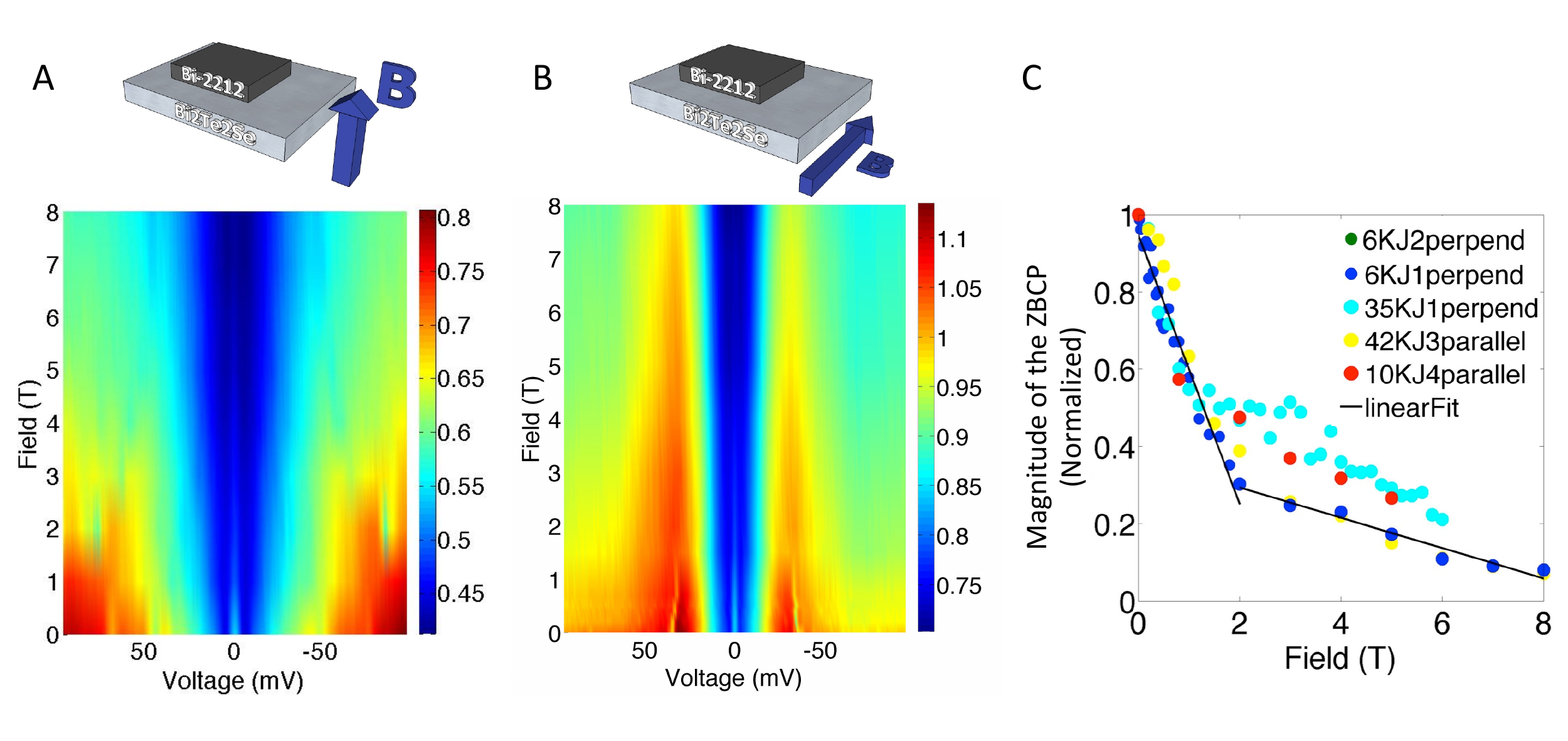}
\caption{\label{fig:mag} (color online) (\textbf{A}) Normalized differential conductance characteristics of J1, at various magnetic fields applied perpendicular to the junction. (Top) Geometry of the junction and the direction of the field. (\textbf{B}) Normalized differential conductance characteristics of J3, at various magnetic fields applied parallel to the junction. (\textbf{C}) Magnitude of the ZBCP for different Bi-2212/Bi$_2$Te$_2$Se junctions versus magnetic field. For clarity, conductances at different fields are divided by the conductance at zero field. The black lines are a guide to the eye.}
\end{figure*}

\begin{figure*}
\includegraphics[width=1\textwidth]{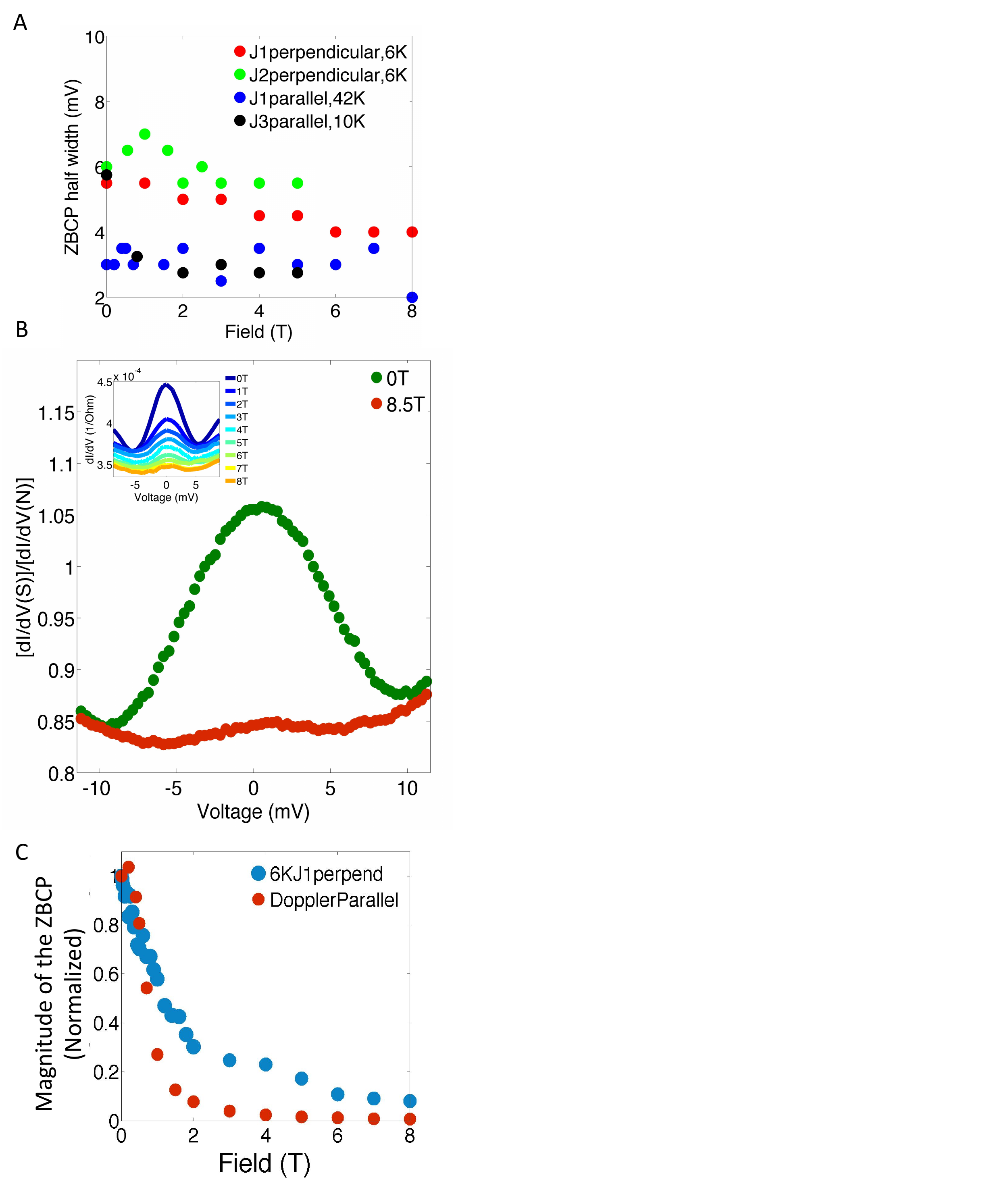}
\caption{\label{fig:width} (color online) (\textbf{A}) 
The half width of the ZBCP in different Bi-2212/Bi$_2$Te$_2$Se junctions, as a function of magnetic field, measured by averaging the minimum of conductance at the positive voltage side and negative voltage side. (\textbf{B}) High-resolution (300 $\mu$V) differential conductance measurement of J2 at 10K normalized by normal-state conductance (110K) for 0 and 8.5 T applied perpendicular to the junction. The inset shows the dI/dV of junction 1 at 6K at various perpendicular applied magnetic fields up to 8 T. (\textbf{C}) The magnitude of the ZBCP in Bi-2212/Bi$_2$Te$_2$Se junction 1 at 6K as a function of parallel magnetic field, compared with the calculation of the Doppler effect suppression of a ZBCP.}
\end{figure*}

Variation of the dI/dV spectrum as a function of magnetic field can help us distinguish between other possible causes of this ZBCP. As discussed later, due to the requirement of enclosed flux, then a ZBCP originating from CAR or WAL must respond anisotropically to applied magnetic field. Furthermore, if the ZBCP originates from the Kondo effect or magnetic impurities, it is expected to be split by the application of field. We explore these possibilities in Fig. \ref{fig:mag}A\&B, which show the differential conductance of two Bi-2212/Bi$_2$Te$_2$Se devices at 10K in magnetic fields applied perpendicular or parallel to the junction interface, respectively. The overall conductance of the spectra decreases, while the conductance at zero-bias is suppressed at a faster rate. The height of the ZBCP goes down identically in parallel and perpendicular applied magnetic fields (see Fig. \ref{fig:mag}C), in contrast to the anisotropic response of the conductance of Bi$_2$Te$_2$Se (resulting from: weak antilocalization (perpendicular field) versus Zeeman shifting of the Dirac cone (parallel field)) (Fig. S1 B) (see Appendix). For fields less than $\sim$ 2 T, the ZBCP height decreases identically for parallel and perpendicular applied field directions, different devices, and temperatures. Furthermore, we do not see splitting/broadening of the ZBCP (see Fig. \ref{fig:width} A \& B), unlike experiments involving ABSs, Kondo effect, or impurities, where a splitting\cite{Deutscher:2005tm} or a broadening and shifting of the ZBCP away from zero is observed\cite{Samokhin:2001ig, 2000Natur.405..764S,Deutscher:2005tm}. In contrast to these experiments, the isotropic suppression of the ZBCP in our data may occur due to the superconducting nodes becoming larger. Furthermore, the ZBCP of various devices all behave the same way, with the only difference being in the crossover field between low and high field slopes. The origin of the crossover is not clear at this time and requires further study.

The magnetic field dependence of our ZBCP further rules out standard Andreev reflection, proximity effect, and ABS as a cause of this peak. Indeed, magnetic fields generate a screening supercurrent resulting in the shifting of the energy of the quasiparticles. This shift is proportional to the dot product of the fermi velocity of the incoming electrons (v$_F$)  and the supercurrent momentum (p$_S$): $\Delta$E = v$_F$.p$_S$ \cite{Tkachov:2005jt}. This so-called \textquotedblleft{Doppler effect}\textquotedblright leads to the reduction of the Andreev peak both in height and width. Specifically, applying the magnetic field parallel to the interface should result in 95\% decrease in the magnitude of the Andreev peak at zero bias in 3T (for g-factor of 2 in the normal material). This decay rate is much higher than the measured decay rate of the ZBCP in our data (Fig. \ref{fig:width}C). Furthermore, while applying magnetic field parallel to the interface creates a shift in the energy of quasiparticles, applying perpendicular magnetic field leads to both negative and positive components of energy shift that average to zero. Therefore, we expect highly anisotropic dependence of the Doppler effect to magnetic field, as previously observed in superconducting proximity devices \cite{Rohlfing:2009ju}, but not in our data. For larger g-factors or perpendicular field direction this rate only increases, which further confirms that the ZBCP in our Bi-2212/Bi$_2$Te$_2$Se junctions is not originating from a simple Andreev reflection. The screening supercurrents have the same effect on ABS. Previous studies have shown that the Doppler effect will cause a suppression of the ABS that is much slower than what we observe in our data and also is highly anisotropic \cite{Deutscher:2005tm,2003PhyC..387..162G}. Therefore the isotropic response of the ZBCP eliminates the possibility that we are tunnelling into the AB-plane of the Bi-2212.

As mentioned earlier Coherent Andreev reflection (reflectionless tunnelling) should also respond to magnetic field anisotropically. CAR results from the constructive interference of multiple scattering events between impurities and the N/Sc interface, and leads to the enhancement of Andreev reflection \cite{1992PhRvL..69..510V,Giazotto:2001dq}. However the application of a voltage or magnetic field results in a phase shift diminishing the constructive interference, ultimately leading to a reduction in the enhancement. When the applied $B$ and $V$ increase, this phase shift naturally leads to a cutoff voltage ($V^*$) and a cutoff magnetic field ($B^*$). As described in the supplemental, from the $V^*$ measured (5.25 meV), we estimate a $B^*$ of 0.4 T, well bellow the field at which the ZBCP is observed to survive ($>$ 8.5 T). Thus we conclude the ZBCP observed between Bi-2212 and Bi$_2$Te$_2$Se can not be due to CAR. Moreover the fact that our ZBCP is reduced at the same rate for perpendicular and parallel fields, further confirms that this peak does not originate from reflectionless tunnelling. Indeed, perpendicular magnetic field results in an enclosed flux in the plane of the TI surface states, but parallel field do not. So we expect to see much faster suppression of the ZBCP in perpendicular magnetic field direction than parallel. The same argument rules out weak antilocalization as a source of our zero-bias peak.

One might argue that if CAR was happening in three dimensions in the bulk Bi$_2$Te$_2$Se rather than the two-dimensional surface states, isotropic suppression of the ZBCP might be observed. However, let us consider the relationship between the cutoff field and effective mass. Specifically assuming ${v_f}^2 = E_f/m$, we obtain (see supplemental):

\begin{eqnarray}
\frac{B^*_{{(c)}}}{B^*_{c_{(AB)}}} = (\frac{m_{(c)}}{m_{(AB)}}) (\frac{{\tau}_{{m}_{(AB)}}}{{\tau}_{{m}_{(c)}}})  (\frac{{\tau}_{{\Phi}_{(AB)}}}{{\tau}_{{\Phi}_{(c)}}})
\end{eqnarray}

where ${\tau}_{\phi}$ and ${\tau}_m$ are the phase-coherence time and momentum relaxation time respectively. Numerous studies have shown the ${\tau}_{\phi}$ and ${\tau}_m$ to be isotropic in the c-axis and AB-plane of Bi$_2$Se$_3$ \cite{Tichy:1979te,Eto2010}. Furthermore, optics and quantum oscillation measurements have shown that the effective mass ratio between the c-axis and the AB-plane varies with carrier density ($m_{(c-axis)}/m_{(AB-plane)} \sim $ 2-8) \cite{Tichy:1979te}. Thus, the cut-off field is expected to be highly anisotropic in the perpendicular and parallel field directions ($B_{c_{(c-axis)}}/B_{c_{(AB-plane)}} \sim 2-8$). This is inconsistent with the isotropic dependence of the ZBCP in response to magnetic field, observed in our data (Fig. \ref{fig:mag}C). We note that disorder in the junction could reduce the AB-plane scattering time, leading to a change in the predicted anisotropy due to CAR. However this would need to perfectly cancel the anisotropy of the effective mass, which seems unlikely to occur perfectly in multiple junctions, as observed here.

We now turn to the possibility of impurities or the Kondo effect \cite{2000Natur.405..764S,Deutscher:2005tm}. ZBCPs originating from either effects are expected to Zeeman split (assuming a g-factor of 2, by 8.5 T we would see (at least) a 980 $\mu$V splitting). In Fig. \ref{fig:width}B, we compare high-resolution (300 $\mu$V) dI/dV scans taken at 0 T and 8.5 T for junction 2 (obtained by thermal cycling of junction 1), which clearly shows that the ZBCP does not split. Shiba states typically arise as finite-bias peaks at zero magnetic field. These states are ABSs emerging as a result of the exchange coupling between impurity states and the superconductor. They move and merge to zero-bias in parallel magnetic fields. \cite{Lee:2012ft} These zero-bias states are inconsistent with our data as well. Majorana fermions can also create a ZBCP. However, no theoretical studies exist for our exact configuration, though the closest work suggests they too should respond anisotropically \cite{Anonymous:H70z94jd}. 

\section{\label{sec:level1}CONCLUSION}

In summary, we have observed a zero-bias conductance peak appearing in multiple Bi-2212/Bi$_2$Te$_2$Se junctions at temperatures below $\sim 0.8T_c$ of Bi-2212. A careful study of the temperature and magnetic field dependence of this ZBCP demonstrates that it is inconsistent with the known effects that can create a ZBCP. Specifically the continuous suppression of the zero-bias conductance below 0.8 T$_c$ rules out Andreev reflection, proximity effect, and ABS. Moreover,  ABS, CAR, WAL, Kondo effect, and magnetic impurities should be strongly sensitive to the orientation of the applied field and/or should result in a splitting of the peak. However none of the above effects were observed. Further studies are needed to shed light on the origin of this zero-bias anomaly.

\section{\label{sec:level1} ACKNOWLEDGEMENTS}

We acknowledge Y. Tanaka, J. Linder, T. Klapwijk, G. Koren, Y. Ran, and Hae-Young Kee for very helpful discussions. The work at the University of Toronto was supported by the Natural Sciences and Engineering Research Council of Canada, the Canadian Foundation for Innovation, and the Ontario Ministry for Innovation. KSB acknowledge support from the National Science Foundation (grant DMR-1410846). The work at Brookhaven National Laboratory (BNL) was supported by DOE under Contract No. DE-AC02-98CH10886. The crystal growth at Princeton was supported by the US National Science Foundation, grant number DMR-0819860.

\bibliography{refe}
\bibliographystyle{apsrev4-1}

\end{document}